\documentclass[10pt,twocolumn]{article}

\usepackage[a4paper,margin=0.75in]{geometry}
\usepackage{amsmath,amssymb,amsfonts}
\usepackage{graphicx}
\usepackage{booktabs}
\usepackage{multirow}
\usepackage{array}
\usepackage{cite}
\usepackage{url}
\usepackage{hyperref}
\usepackage{caption}
\usepackage{subcaption}
\usepackage{algorithm}
\usepackage{algpseudocode}
\usepackage{xcolor}
\usepackage{xurl}

\hypersetup{
    colorlinks=true,
    linkcolor=blue,
    citecolor=blue,
    urlcolor=blue
}

\title{\textbf{Solar Cars: A Comprehensive Review}}

\author{
    Afsaneh Mollasalehi$^{1}$,
    Armin Farhadi$^{2}$ \\
    \small $^{1}$Faculty of Technology, Natural Sciences and Maritime Sciences, \\
    \small University of South Eastern Norway, Kongsberg, Norway \\
    \small $^{2}$School of Electrical and Computer Engineering, \\
    \small College of Engineering, University of Tehran, Iran
}

\date{}

\begin{document}

\maketitle

\begin{abstract}
Energy crisis has forced many countries to think of a replacement for energy supply. Renewable energy sources as friendly environment sources play a pivotal role in producing clean energy for various sectors in industry. Gas emissions originating from the transportation industry are another contributing factor to air pollution. Hence, designing and utilizing vehicles that run on renewable energy is crucial, as it provides a dependable energy source that is naturally abundant, leaves nearly no carbon footprint, and is sustainable. Solar powered electric cars make a significant impact on global climate change. To better understand this impact and building upon the plenty of research done on this topic, this paper aims to provide a comprehensive review of the various factors related to solar cars. Specifically, this review examines the following key factors: types and sizing of solar cars, solar vehicle power source configurations, leading solar car nations, and solar car challenges.
\end{abstract}

\noindent\textbf{Keywords:} 
Solar Cars, Solar Vehicles, Electric Vehicles, Solar Energy, Solar Panels

\section{Introduction}

The world is shifting toward environmentally friendly energy sources. Various factors are behind this global trend including the depletion of conventional fossil fuel sources, population growth, technological advancement, and other related drivers. Renewable Energy Sources (RESs) promise clean alternative solutions by replacing fossil fuels with renewables such as solar and wind, which significantly reduce Greenhouse Gas Emissions (GHG)~\cite{ref1}.

\begin{figure}[!t]
    \centering
    \includegraphics[width=0.95\columnwidth]{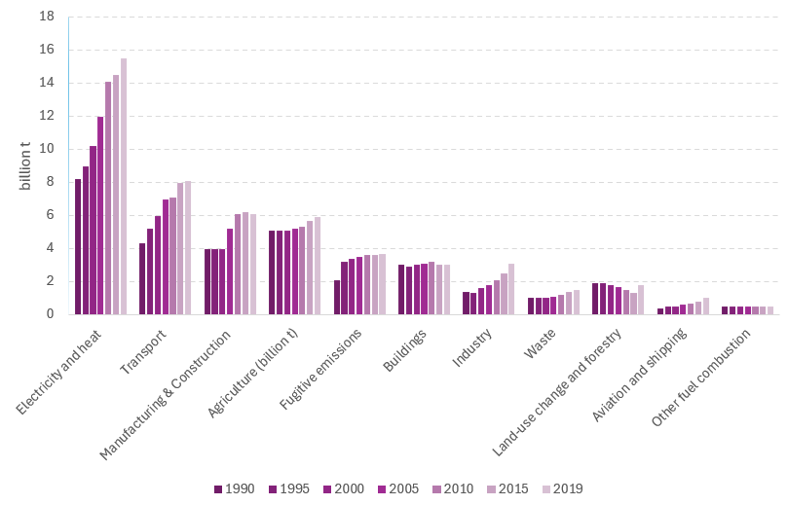}
    \caption{The total amount of greenhouse gases emitted globally each year, separated by sector and expressed in tonnes of CO$_2$-equivalent~\cite{ref2}.}
    \label{fig:ghg}
\end{figure}

All industrial sectors contribute to environmental pollution to varying degrees. However, a substantial portion of pollution originates from the transportation sector. As shown in Fig.~\ref{fig:ghg}, the transport sector accounts for a significant portion of global GHG emissions. Moreover, according to the Environmental Protection Agency (EPA), transportation accounted for 26\% of global GHG emissions in 2014. Hence, Electric Vehicles (EVs) have been proposed to reduce GHG emissions and provide a more sustainable solution. In this context, Solar Energy (SE) is considered a low-emission technology that produces electricity through solar modules integrated into the vehicle body. Solar Cars (SCs), as a subset of Solar Vehicles (SVs), are designed to run entirely on power derived from the sun~\cite{ref2}.

SCs are not a recent development; in fact, their history dates back to the mid-20th century, beginning with small-scale models. In 1955, William G.~Cobbs of General Motors created a 15-inch solar-powered model called the \textit{Sunmobile}, which used 12 selenium photovoltaic cells and a small electric motor.

This was followed in 1958 by the International Rectifier Company, which produced the first Solar Car (SC) capable of carrying a human driver, a converted 1912 Baker electric car fitted with more than 10,000 solar cells. Throughout the 1970s, progress continued. In 1977, Professor Ed Passerini at Alabama University built the \textit{Bluebird}, a full-scale prototype designed to run solely on photovoltaic power without a battery.

Between 1977 and 1980, several innovations emerged. Professor Masaharu Fujita in Japan created a four-wheeled SC from two solar bicycles. In 1979, English inventor Alain Freeman built a three-wheeled SC. In 1980, Arye Braunstein and colleagues at Tel Aviv University developed a SC with Solar Panels (SPs) on the hood and roof, capable of generating 400 watts of peak power.

A major milestone occurred in 1982, when Hans Tholstrup and Larry Perkins completed the first intercontinental journey in a solar-powered car, traveling from Perth to Sydney, Australia. Tholstrup later founded the \textit{World Solar Challenge} in Australia. Subsequent records include the \textit{Sunrunner} in 1984, which reached 24.7 mph with a battery, and a 1986 vehicle that achieved 41 mph without a battery, setting a Guinness World Record. In 1987, \textit{GM Sunraycer} completed a 1,866-mile trip at an average speed of 42 mph~\cite{ref3}.

Since then, SC development has continued mainly within universities and competitions. The 21st century has seen growing interest in solar hybrids from automakers like Ford and Mazda, integrating SPs into vehicle systems to support electrification, demonstrating that SCs are not a recent development, but rather the product of decades of sustained engineering effort. With the energy crisis, the attention to SCs is still ongoing, and it has become more prominent~\cite{ref3}.

However, designing and utilizing such vehicles presents several challenges. Despite significant research into Photovoltaic (PV) technology, some specific issues still pose relatively unique barriers~\cite{ref4}. Moreover, the market, particularly the automotive industry, is not sufficiently prepared for the widespread practical adoption of this rapidly advancing technology~\cite{ref4}. The path to widespread adoption is slowed down by a series of challenges discussed in the following paragraphs. This review primarily uses recent articles (from 2020 onwards) relevant to the title. However, articles from earlier years have also been used to a limited extent, only if they contain relevant information. This paper is structured as follows. Subsequent sections address the types and sizing of SCs, power source configurations, leading SC nations, and the key challenges involved in their development.

\section{Types and Sizing of SCs}

\begin{figure}[!t]
    \centering

    \begin{subfigure}{0.48\columnwidth}
        \centering
        \includegraphics[width=\linewidth]{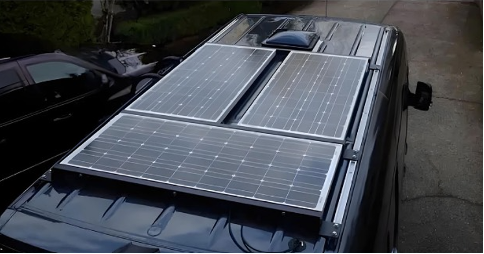}
        \caption{VAPV configuration.}
        \label{fig:vapv1}
    \end{subfigure}
    \hfill
    \begin{subfigure}{0.48\columnwidth}
        \centering
        \includegraphics[width=\linewidth]{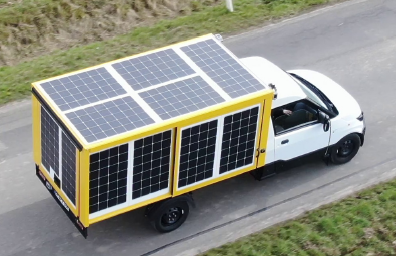}
        \caption{VIPV integration.}
        \label{fig:vapv2}
    \end{subfigure}

    \caption{Comparison of VAPV and VIPV installations on solar vehicles~\cite{ref2}.}
    \label{fig:vapv}
\end{figure}

SCs have been considered for various purposes. When it comes to commercial use cases, several manufacturers have introduced commercial Solar Electric Vehicles (SEVs). In some cases, PV is mounted onto the existing body of the vehicle. This is referred to as Vehicle-Added PV (VAPV), where the panels are attached to surfaces such as the roof without replacing any structural parts, as shown in Fig.~\ref{fig:vapv}, and serve only as an energy source. 

\begin{figure}[!t]
    \centering
    \includegraphics[width=0.95\columnwidth]{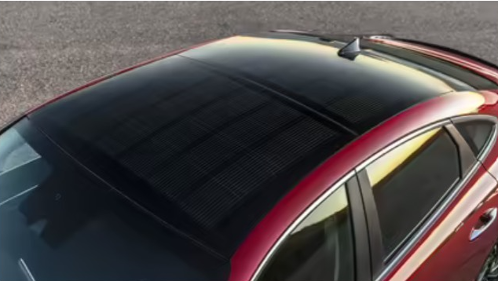}
    \caption{Hyundai Sonata Hybrid with curved PV surfaces.}
    \label{fig:vipv}
\end{figure}

In other cases, PV is designed as an integral part of the vehicle body, replacing structural components such as the roof, hood, or windows, as shown in Fig.~\ref{fig:vipv}. This is called Vehicle-Integrated PV (VIPV), where the panels function both as body elements and energy generators, offering better integration and aesthetics~\cite{ref2}.

Established automotive manufacturers such as Toyota and Hyundai have already introduced VIPV options for their vehicles. The Toyota Prius offers a 180~Wp (watt-peak: maximum solar panel power under standard sunlight conditions) PV panel that charges the traction battery while parking and the auxiliary battery while driving. Hyundai has also released the IONIQ~5 with a solar roof option~\cite{ref5,ref6}.

\begin{figure}[!t]
    \centering
    \includegraphics[width=0.95\columnwidth]{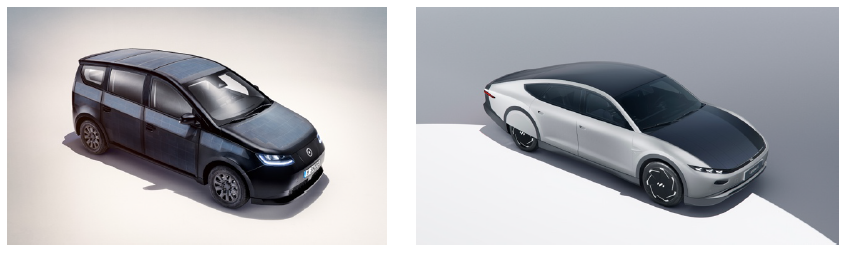}
    \caption{Sono Sion (left, image provided by Sono Motors GmbH) alongside the Lightyear~0 (right, image provided by Lightyear)~\cite{ref2}.}
    \label{fig:sion_lightyear}
\end{figure}

Among startup companies, Sono Motors, Lightyear, and Aptera have showcased purpose-built VIPV products. Sono Motors announced the Sion, shown in Fig.~\ref{fig:sion_lightyear}, a family van equipped with 456 half-cut cells (1200~Wp VIPV system) capable of directly charging a 54~kWh traction battery. The Sion adds 112~km per week to its 305~km total range from its 248 integrated solar cells. Sono Motors also offers after-market PV products for commercial vehicles such as vans, trucks, and buses.

Lightyear introduced the \textit{Lightyear 0}, shown in Fig.~\ref{fig:sion_lightyear}, a luxury sedan featuring a 1005~Wp VIPV system with IBC cells and a 60~kWh battery. It provides a combined range of 725~km and can add 7,000 to 20,000~km annually from solar energy alone.

Aptera has announced a highly aerodynamic car with a 700~Wp VIPV charging system, generating up to 700~watts of solar power and offering a daily range of 40~miles. Other notable SVs include Clean Motion, a small delivery van that can add 130~km per day under bright sunlight, and Squad, a micro car with a solar range of about 20~km. Given that the average micro car usage in Europe is 12~km per day, Squad can operate as a fully sustainable vehicle for daily urban mobility. Notably, Sono Motors and Lightyear have announced a transition from producing entire vehicles with VIPV toward selling PV kits for vehicles~\cite{ref6,ref7,ref8,ref9,ref10}.

\section{SC Power Source Configurations}

As the name suggests, SCs are expected to work solely on energy derived from the sun. However, they cannot yet be considered a practical replacement for conventional cars. The main obstacles include maximum power output, limited driving range, dimensions, and high costs. This is why other types of SCs, including Hybrid Solar Vehicles (HSVs), which operate with supplementary energy sources, have been proposed.

HSVs integrate photovoltaic panels into Hybrid Electric Vehicles (HEVs) to combine the benefits of both technologies. Unlike conventional HEVs that use charge-sustaining strategies (keeping the battery State of Charge (SOC) at a constant level throughout driving), HSVs require charge-depletion strategies. This means the battery is intentionally discharged during driving so that it has enough empty capacity to store solar energy captured by the PV panels during the subsequent parking phase.

Therefore, at the end of a driving trip, the final SOC must be low enough to allow full storage of the solar energy expected during the next parking period. However, maintaining an unnecessarily low final SOC would lead to two major problems: first, additional energy losses due to deeper discharging, and second, a significant reduction in battery lifetime.

To solve this conflict, the optimal management of the battery requires prior knowledge of how much solar energy will be available during the next parking phase. This can be achieved through real-time access to weather forecasts, allowing the vehicle's energy management system to determine the ideal final SOC for each driving trip~\cite{ref11}.

In fact, advances in battery technology have been a key driver behind the growing popularity and widespread adoption of electric vehicles. In 1996, the Lithium Iron Phosphate (LFP) battery was invented by John~B.~Goodenough. The LFP battery is a type of lithium-ion battery with a cathode made of lithium iron phosphate and a graphite-based anode. It is widely used in electric vehicles such as cars, buses, and trucks.

The key specifications include an output voltage of 72~V, a current capacity of 25~ampere-hours, a power rating of 1800~watts, and a configuration of 90~cells with each cell providing 3.2~V and 3.27~A. The battery offers an impressive lifespan of up to 12,000~cycles, depending on the load. Charging time varies from approximately 13--14~hours via AC charging to only about 3~hours when using SPs.

Another important configuration related to batteries is the Smart Battery Management System (BMS), which is essential for monitoring, reporting, and maintaining battery health and performance. The BMS is equipped with Bluetooth-enabled monitoring and remote on/off capabilities. It protects individual cells from damage, noting that overcharging has minimal impact on capacity, whereas over-discharge significantly reduces both capacity and lifespan.

In extreme cases, either condition can cause irreversible damage. Therefore, it is recommended to gradually use the battery's capacity to extend its operational life and ensure long-term peak performance~\cite{ref12}.

\section{Leading SC Nations}

Several countries have conducted and continue to conduct research on SCs. The fundamental motivations behind the research interest in countries such as the United States, Germany, and France are significant. Reducing the use of fossil-fuel-powered vehicles and increasing electric car adoption due to diminishing traditional fuel supplies could be among the main reasons~\cite{ref13}.

The fact that the United States and Australia host major solar competitions, including Australia’s 3000-km \textit{World Solar Challenge} and the \textit{American Solar Challenge} from Omaha to Bend, Oregon, demonstrates the growing importance of SCs and proves that these countries are making substantial progress in this field~\cite{ref11,ref14}.

Similarly, India has hosted the \textit{Shell Eco-marathon Urban Concept Battery Electric} competition addressing urban electric vehicles, indicating that the country is also paying attention to this area~\cite{ref15}. Despite energy constraints, Pakistan is actively progressing in the field of solar vehicle adoption as well. One experiment involved the design of a micro solar-powered vehicle, where HOMER Pro was employed to determine the appropriate system sizing~\cite{ref14}.

\section{SC Challenges}

SE has gained significant research attention due to its potential as a power source. However, in the past, SPs were not efficient enough to be considered a practical alternative to fossil fuels. The main problem was their inability to capture and convert sunlight into usable energy effectively enough to compete with conventional energy sources. This inefficiency made them unsuitable for large-scale energy production or for replacing fossil fuels in applications such as transportation and electricity generation.

The turning point came with advancements in the materials used to manufacture SPs. These technological improvements led to a significant increase in SP efficiency, meaning they could convert a higher percentage of sunlight into electricity. As a result, SPs became more practical and effective for energy production~\cite{ref16}.

Another key challenge in solar vehicle design is the performance loss caused by curved PV surfaces, as shown previously in Fig.~\ref{fig:vipv}. Unlike flat panels installed on rooftops, as shown in Fig.~\ref{fig:vapv}, PV cells integrated into a car body, such as on the hood, roof, and rear roof, inevitably follow the curved shape of the vehicle.

Experimental road tests using high-efficiency 34\% multi-junction solar cells revealed that curved PV modules suffer significant power losses, with performance ratios varying between 0.3 and 1.1 even under favorable sun heights (35$^\circ$--75$^\circ$). These losses are primarily attributed to cosine loss (non-perpendicular sunlight incidence) and self-shading (one part of the curved panel casting shadows on another).

Although simulations and stationary tests had previously indicated such losses, real-driving experiments validated their severity. The proposed surface model, which accounts for curvature-related losses, can also be extended to estimate energy yield across different climates, assuming random vehicle orientation~\cite{ref4}.

The challenges related to chassis design in solar-powered electric vehicles are addressed in~\cite{ref12}. Engineers must carefully balance material selection, structural integrity, and weight optimization. Positioning the battery without compromising space utilization or safety standards is another critical issue. Furthermore, achieving aerodynamic efficiency while accommodating SPs, ensuring manufacturability, maintaining cost-effectiveness, and properly integrating the suspension system all add layers of complexity to the chassis design~\cite{ref12}.

To provide a clear overview of the existing literature, Table~\ref{tab:literature_review} summarizes the main studies cited in this paper. For each reference, the table presents the type of solar vehicle, the energy source configuration, and the key findings or challenges discussed. This summary highlights the current research landscape and the gaps addressed in the present work.

\begin{table*}[!t]
\centering
\caption{Overview of Selected Literature in This Paper}
\label{tab:literature_review}

\small
\setlength{\tabcolsep}{4pt}

\begin{tabular}{|c|p{2.3cm}|p{3.2cm}|p{2cm}|p{5.2cm}|}
\hline
\textbf{Ref} &
\textbf{Types / Sizing} &
\textbf{Power Source Configurations} &
\textbf{Countries} &
\textbf{Challenges} \\
\hline

\cite{ref2}, \cite{ref5}, \cite{ref6}, \cite{ref3}, \cite{ref7}, \cite{ref8}, \cite{ref9}, \cite{ref10}
&
VIPV and VAPV; Toyota Prius, Hyundai IONIQ 5, Sono Sion, Lightyear 0, Aptera
&
Battery storage, direct traction battery charging via VIPV, DC-DC converter
&
Germany, U.S., Netherlands, Japan, Korea
&
Curved surface power loss, dynamic shading, temperature effects, dual testing standards, energy modeling uncertainty
\\
\hline

\cite{ref11}
&
Pure solar cars and HSVs; HSV prototype based on Piaggio Porter
&
Pure SCs: only solar energy
&
Not directly mentioned
&
Limited power output, battery degradation due to deep discharging, need for weather forecast-based energy management
\\
\hline

\cite{ref12}
&
Urban EV, 4-wheeler, 4-seater
&
Solar PV panels + charge controller + LFP battery + PM BLDC motor + DC-DC converters + AC charging
&
India
&
Lightweight design, aerodynamic optimization, fabrication feasibility, battery management, MPPT limitations, high EV cost
\\
\hline

\cite{ref14}
&
Micro solar electric vehicle, commercial SCs, solar race cars
&
Solar PV panels + battery storage + DC motor
&
Pakistan
&
Energy crisis, low income, lack of charging infrastructure, limited PV integration space, high EV prices
\\
\hline

\cite{ref4}
&
Toyota Prius PHEV (VIPV)
&
PV modules + battery + drivetrain
&
Japan
&
Curved surface loss, cosine loss, self-shading, dynamic shading, irradiance fluctuation, temperature variation
\\
\hline

\cite{ref13}
&
Solar-powered electric vehicle (SPEV)
&
Solar panels + AC charging station + battery pack + inverter + motor controller
&
U.S., Germany, France
&
Slow solar charging, limited roof area, bad weather dependency, battery weight issues
\\
\hline

\end{tabular}
\end{table*}

\section{Conclusion}

Increasing energy demand and the need to reduce GHG emissions caused by population growth have paved the way for adopting efficient and sustainable energy solutions. In response to population growth, various industries continue to expand, contributing significantly to carbon emissions as well as the depletion of non-renewable fuel resources. Among all industries, the transportation sector is a major contributor to climate change, along with growing concerns regarding environmental pollution and fossil fuel depletion. Hence, utilizing environmentally friendly vehicles that operate using electricity derived from solar energy presents a promising solution. Many countries have pursued solar-powered electric cars through competitions and real-world implementations by automobile manufacturers. Moreover, ongoing advancements in solar cell efficiency and battery storage capacity have played a crucial role in supporting this clean and sustainable transportation option. Given these advancements, this paper attempted to cover the fundamental and essential aspects of solar-powered electric vehicles to provide a general overview of the field. Important issues such as passenger capacity, battery storage, and leading countries active in this area were discussed. Other critical aspects, including PV surface types and their associated challenges, were also addressed.
For future studies, more attention should be paid to the real-world energy yield of curved PV surfaces under different driving and parking conditions. In addition, future research could focus on low-cost solar EV designs for developing countries, where charging infrastructure is limited but solar radiation is often abundant.



\begin{thebibliography}{99}

\bibitem{ref1}
A. Mollasalehi and A. Farhadi, ``Solar and wind power forecasting: A comparative review of LSTM, random forest, and XGBoost models,'' \emph{arXiv preprint arXiv:2509.24059}, 2025.

\bibitem{ref2}
N. R. Patel, ``Analysis of the performance of vehicle-integrated photovoltaics: Solar photovoltaics on the move,'' 2025.

\bibitem{ref3}
P. O. Babalola and O. E. Atiba, ``Solar powered cars -- A review,'' \emph{IOP Conference Series: Materials Science and Engineering}, vol. 1107, no. 1, p. 012058, 2021.

\bibitem{ref4}
K. Araki \emph{et al.}, ``Rating vehicle-integrated photovoltaics: Power and energy loss by curved surface,'' \emph{Solar Energy Materials and Solar Cells}, vol. 292, p. 113814, 2025.

\bibitem{ref5}
F. Lambert, ``Toyota brings back the solar panel on the Plug-In Prius Prime -- but now it powers the car,'' \emph{Electrek}, 2016.

\bibitem{ref6}
J. McCann, ``Hyundai’s new electric car has a solar panel roof and can charge other EVs,'' \emph{TechRadar}, 2021.

\bibitem{ref7}
``Sono Motors,'' Available: \url{https://sonomotors.com/}

\bibitem{ref8}
``Lightyear,'' Available: \url{https://lightyear.one/lightyear-0}

\bibitem{ref9}
``Lightyear transition,'' Available: \url{https://lightyear.one/press-releases/lightyear-announces-leadershipchange}

\bibitem{ref10}
``Sono Sion program termination,'' Available: \url{https://sonomotors.com/en/press/press-releases/sonomotors-commits-to-focus-exclusively-on-solar-tech-company-and-has-terminated-the-sion-program/}

\bibitem{ref11}
G. Rizzo, I. Arsie, and M. Sorrentino, \emph{Hybrid Solar Vehicles}. INTECH Open Access Publisher, 2010.

\bibitem{ref12}
J. Heeraman, R. Kalyani, and B. Amala, ``Towards a sustainable future: Design and fabrication of a solar-powered electric vehicle,'' \emph{IOP Conference Series: Earth and Environmental Science}, vol. 1285, no. 1, p. 012035, 2024.

\bibitem{ref13}
F. Mohammadi, ``Design, analysis, and electrification of a solar-powered electric vehicle,'' \emph{Journal of Solar Energy Research}, vol. 3, no. 4, pp. 293--299, 2018.

\bibitem{ref14}
A. Husnain and M. T. Iqbal, ``System design and PV sizing of a micro solar electric vehicle for Pakistan,'' in \emph{2022 IEEE Conference on Technologies for Sustainability (SusTech)}, pp. 65--70, 2022.

\bibitem{ref15}
M. H. Ahmadi \emph{et al.}, ``Evaluation of electrical efficiency of photovoltaic thermal solar collector,'' \emph{Engineering Applications of Computational Fluid Mechanics}, vol. 14, no. 1, pp. 545--565, 2020.

\bibitem{ref16}
A. H. Elsheikh \emph{et al.}, ``Low-cost bilayered structure for improving the performance of solar stills,'' \emph{Sustainable Energy Technologies and Assessments}, vol. 49, p. 101783, 2022.

\end{thebibliography}
\end{document}